\documentclass[letter,traditabstract]{aa} % for the letters
\usepackage{graphicx}
\usepackage{txfonts}
\usepackage[rightcaption]{sidecap}
\usepackage{natbib}
\bibpunct{}{}{;}{a}{}{,}

\begin{document}

\title{Discovery of a massive X-ray luminous galaxy cluster at $z$=1.579
\thanks{
This work is based on observations obtained with XMM-Newton, an ESA science mission 
with instruments and contributions directly funded by ESA Member States and the USA (NASA);
and on observations carried out using the New Technology 
Telescope and the Very Large Telescope  at the La Silla Paranal Observatory, 
under Program IDs  079.A-0634(B), 081.A-0312(A) and 084.A-0844(A).}}

   \author{J.S. Santos \inst{1},
            R. Fassbender  \inst{2},
            A. Nastasi  \inst{2},
             H. B\"ohringer \inst{2},
              P. Rosati \inst{3},           
             R. \v{S}uhada \inst{2},
             D. Pierini \inst{2}\thanks{guest astronomer at MPE},
             M. Nonino \inst{4},
            M. M\"uhlegger \inst{2},
          H. Quintana \inst{5},
          A.D. Schwope \inst{6},
          G. Lamer \inst{6},
          A. de Hoon \inst{6},
          V. Strazzullo  \inst{7}
	}
   \institute{
	     \inst{1} European Space Astronomy Centre (ESAC)/ESA, Madrid, Spain \\
            \email{jsantos@sciops.esa.int} \\
	     \inst{2} Max-Planck-Institut f\"ur extraterrestrische Physik,  Giessenbachstra\ss e, 85748 Garching, Germany \\
            \inst{3} European Southern Observatory, Karl-Schwarzschild Strasse 2, 85748 Garching, Germany \\
             \inst{4}  INAF-Osservatorio Astronomico di Trieste, Via Tiepolo 11, 34133 Trieste, Italia  \\
            \inst{5}    Departamento de Astronom\'ia y Astrof\'isica, Pontificia Universidad Cat\'olica de Chile, Casilla 306, Santiago 22, Chile \\
            \inst{6}   Astrophysikalisches Institut Potsdam (AIP), An der Sternwarte 16, 14482 Potsdam, Germany \\
            \inst{7}   CEA \/ Saclay, Service d'Astrophysique, L'Orme des Merisiers, B\^at. 709, 91191 Gif-sur-Yvette Cedex, France \\
             }

   \date{Received ... ; accepted ...}

 \abstract
{We report on the discovery of a very distant galaxy cluster serendipitously detected in the
archive of the XMM-\textit{Newton} mission, within the scope of the XMM-Newton Distant Cluster Project (XDCP). 
XMMUJ0044.0-2033 was detected at a high significance level (5$\sigma$) as a compact, but significantly extended 
source in the X-ray data,  with a soft-band flux $f(r<40\arcsec)$=(1.5$\pm$0.3)$\times$10$^{-14}$erg s$^{-1}$cm$^{2}$.

Optical/NIR follow-up observations confirmed the presence of an overdensity of red galaxies 
matching the X-ray emission. 
The cluster was spectroscopically confirmed to be at $z$=1.579 using ground-based VLT/FORS2 spectroscopy.  
The analysis of the I-H colour-magnitude diagram shows a sequence of red galaxies with a colour range 
[3.7 $<$ I-H $<$ 4.6] within 1$\arcmin$ from the cluster X-ray emission peak. 
However, the three spectroscopic members (all with complex morphology) have significantly bluer colours relative to 
the observed red-sequence. In addition, two of the three cluster members have [OII] emission, indicative of on-going 
star formation.

Using the spectroscopic redshift we estimated the X-ray bolometric luminosity, $L_{bol,40\arcsec}$ $\sim 5.8\times$10$^{44}$ erg s$^{-1}$,
 implying a massive galaxy cluster.
 This places XMMU J0044.0-2033 at the forefront of massive distant clusters, closing the gap 
 between lower redshift systems and recently discovered proto- and low-mass clusters at $z>$1.6. }

   \authorrunning{J.S.Santos et al.}
   \titlerunning{Discovery of a massive galaxy cluster at z=1.579}

   \maketitle

\section{Introduction}

In the hierarchical clustering scenario for structure formation, galaxy clusters result 
from the gravitational collapse of the densest peaks in the primordial fluctuations, 
and grow through accretion and mergers with neighboring poor clusters or groups. 
(e.g. Colberg et al. \cite{colberg}).
The most massive galaxy clusters are therefore the last structures to form and virialize. 

Fully assembled clusters strongly emit thermal X-rays arising from the intracluster medium (ICM), 
a hot, diffuse plasma that contains most of the cluster baryons. X-ray data is therefore essential to 
assess the dynamical state of a high-redshift overdensity of galaxies, allowing us to discriminate 
between a cluster and a proto-cluster. 
During the last decade, we have witnessed a tremendous improvement on the detection
and study of massive clusters up to $z=$1 and beyond, thanks to the \textit{Chandra} and XMM-\textit{Newton}
observatories. While clear X-ray emission has been detected in several massive $z >$1.4 clusters,
(Stanford et al. \cite{stanford}, Fassbender et al. \cite{fassben11}, Nastasi et al. subm.), the IR selected clusters at 
$z$=1.6 (Papovich et al. \cite{papovich}) and 2.1 (Gobat et al. \cite{gobat}) show no prominent X-ray diffuse emission, 
which prevents the assessment of their dynamical status from an X-ray analysis.

The discovery and study of very distant massive clusters is essential to trace the epoch of cluster 
formation and has important consequences for cosmology. 
Recent observational (e.g. Jee et al. \cite{jee}) and theoretical studies (e.g. Baldi \& Pettorino \cite{baldi}) 
have shown that high-$z$ clusters are important probes to test the standard $\Lambda$CDM  model, either by 
detecting departures from the standard model (e.g. as a test for dynamical coupled dark energy)
or as an indication that the initial conditions are not completely Gaussian (Jimenez \& Verde \cite{jimenez}, 
Sartoris et al. \cite{sartoris}).

Serendipitous searches of the XMM-\textit{Newton} archive provide one of the most efficient ways to find massive 
distant clusters, due to the large area and photon collecting power of this mission.
In particular, the XMM-Newton Distant Cluster Project (XDCP) with a survey area of $\sim$80 deg$^{2}$ and an average 
soft band sensitivity of 0.8$\times$10$^{-14}$ erg s$^{-1}$cm$^{-2}$, has proven to be a successful high-$z$ cluster 
survey, with about 30 clusters confirmed to be at $0.8<z<1.6$ (e.g. Mullis et al. \cite{mullis}, Santos et al. \cite{joana}, 
Fassbender et al. \cite{fassben11}). The  technical details of this program  
can be found in Fassbender et al. 2011 (in prep.).

In this letter we present the properties of a newly discovered galaxy cluster at $z$=1.58, 
using a multi-wavelength dataset covering X-rays, optical/NIR imaging, and 
optical spectroscopy. 
The adopted cosmological parameters are H$_{0}$ = 70 km s$^{-1}$ Mpc$^{-1}$, $\Omega_{m}$ = 0.3, 
$\Omega_{\Lambda}$ = 0.7. In this cosmology, 1$\arcmin$ on the sky corresponds to 508 kpc at $z$ = 1.58.
Filter magnitudes are presented in the Vega system unless stated otherwise.

\begin{figure*}
\sidecaption
\centering
\includegraphics[height=6.8cm,angle=0]{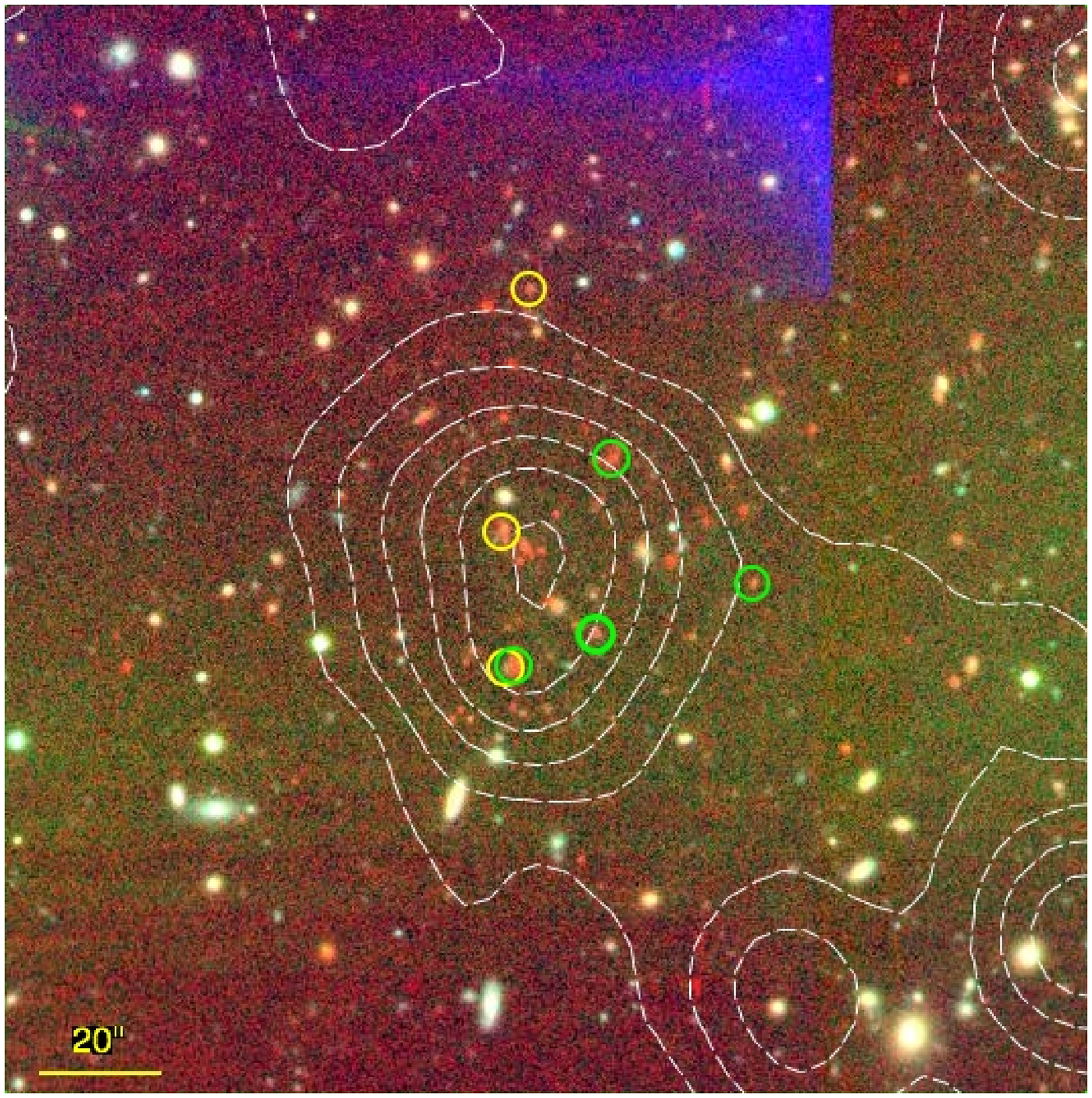}
\includegraphics[height=6.8cm,angle=0]{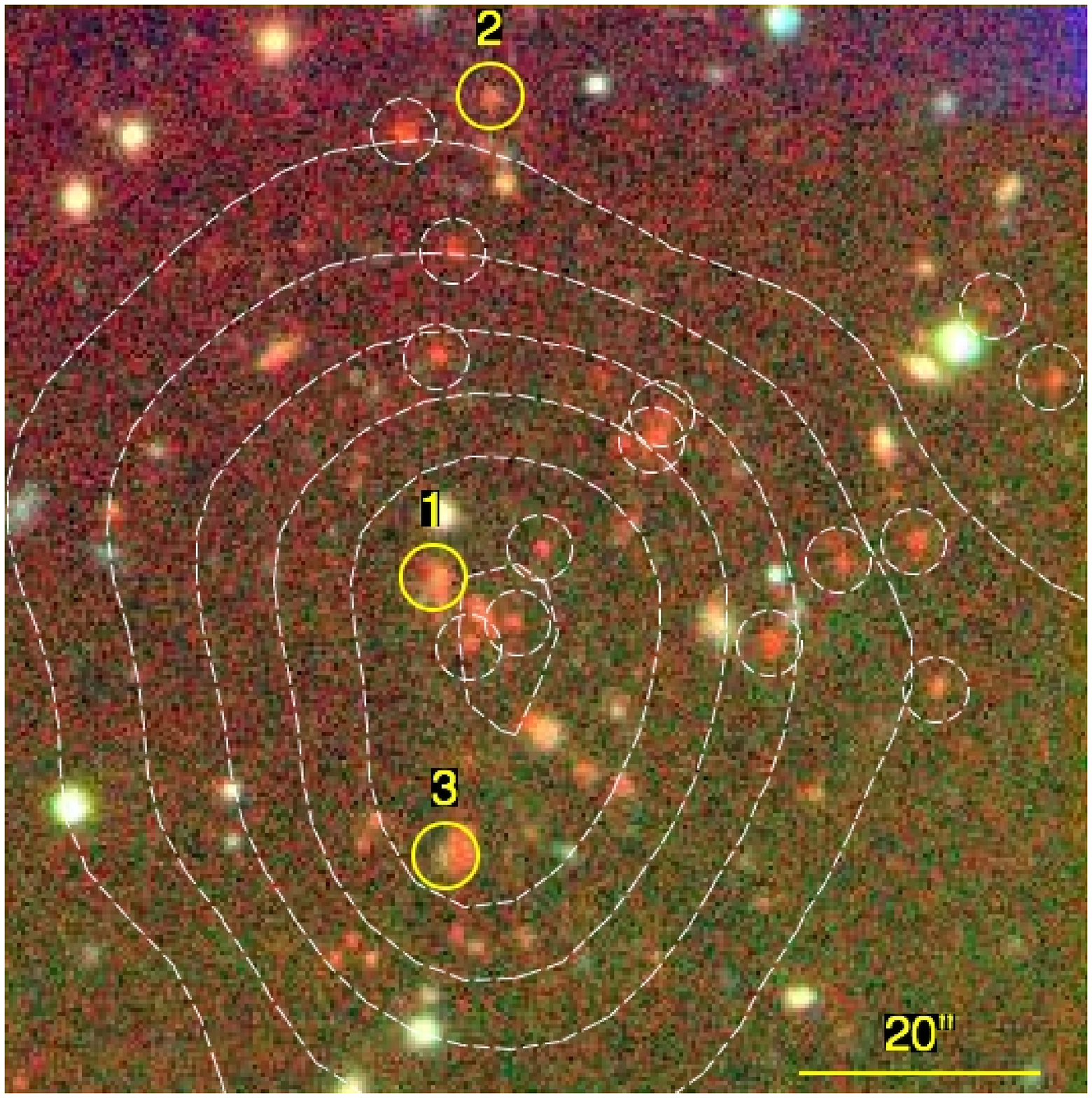}
\caption{IzH colour image of XMMUJ0044 with X-ray contours overlayed in white 
and the 3 spectroscopically confirmed cluster members shown with yellow circles (ID numbers refer to Table 1). 
\textit{Left} Cluster field image (3$\arcmin \times$3$\arcmin$) showing the 4 interlopers (green circles) within 1$\arcmin$.
\textit{Right} Blow-up of the core region (1.5$\arcmin \times$1.5$\arcmin$) showing the red-sequence galaxies
(white circles) - see Fig. 2. North is up and East is to the left. The XMM-\textit{Newton} PSF at the cluster 
off-axis angle is $\sim$10$\arcsec$ (FWHM).}
 \label{lowz}
\end{figure*}

\section{Observations and data analysis}

\subsection{X-ray properties}

XMMUJ0044.0-2033\footnote{Also listed as extended in the 2XMM catalogue
as 2XMM J004405.2-203359 (Watson et al. \cite{watson}).} (hereafter XMMUJ0044, 
RA: 00h44m05.2s, DEC: -20d33m59.7s) was serendipitously 
detected in the XMM-\textit{Newton} field with OBSID=0042340201, at an off-axis angle of 10.8 arcmin.
The nominal exposure time of this observation is 15 ksec, with a clean effective time of 8.5 ksec.
The data reduction and source extraction were performed with {\tt SAS v6.5}.
The cluster candidate was detected as a strong extended source, with an extent significance of 5$\sigma$.
The X-ray emission is compact, however, it is significantly extended compared to the local PSF.

Based on the X-ray contours and lack of an optical counterpart on DSS, XMMUJ0044 was classified as a 
strong distant cluster candidate.
Using the growth curve method we 
measured the cluster unabsorbed soft-band flux (Galactic $N_{H}$=1.9$\times$10$^{20}$cm$^{-2}$) 
within a circular region of 40$\arcsec$radius, $f_{x}$[0.5-2.0]=(1.5$\pm$0.3)$\times$10$^{-14}$ erg/s/cm$^{2}$, 
and within the estimated $R_{500}$, $f_{x}$[0.5-2.0]=(1.6$\pm$0.3)$\times$10$^{-14}$ erg/s/cm$^{2}$.
The scaled radius $R_{500}$ (519 kpc) was obtained using the scaling relations in Pratt et al. (\cite{pratt}).
Using the cluster redshift (see section 2.3) we measure the cluster soft-band luminosity 
$L_{X,40\arcsec}$[0.5-2.0]=(1.8$\pm$0.4)$\times$10$^{44}$ erg/s and bolometric luminosity, 
$L_{bol,40\arcsec}$=5.8$\times$10$^{44}$ erg/s ($L_{bol, R_{500}}$=6.1$\times$10$^{44}$ erg/s).

We followed two approaches to estimate the cluster temperature and mass, in order to evaluate 
statistical and systematic errors associated with these measurements.
In an initial, more empirical approach, we scaled the X-ray properties of another 
XDCP distant cluster, XMMUJ2235.3-2557 at $z$=1.393 (Mullis et al. \cite{mullis}), which were accurately 
measured using high-quality \textit{Chandra} data: $T=8.6^{+1.3}_{-1.2}$ keV and 
$L_{bol,40\arcsec}$=8.5$\times$10$^{44}$ erg/s (Rosati et al. \cite{rosati}). 
The cluster total mass, obtained from weak lensing measurements, is equal to
7.3$\pm$1.3$ \times$10$^{14}$ M$_{\odot}$ (Jee et al. \cite{jee}). 
Applying the empirical scaling relation, $L \propto T^{2.88}$ (Arnaud \& Evrard \cite{arnaud}), and  the self similar 
relation $M \propto T^{3/2}$ (Rosati, Borgani \& Norman \cite{rosati02}), we obtain estimates of the 
temperature and total mass of XMMUJ0044: $T_{est} \approx$ 6.9 keV, $M_{tot} \approx$5$\times$10$^{14}$ M$_{\sun}$.

Following the assumption of a non-evolving $L_{X}-T$ scaling relation and the related slower evolution of the 
 $L_{X}-M$ relation (Fassbender et al. \cite{fass1230}), we estimate an ICM temperature of 
 $T_{est}$=4.8$\pm$1.2 keV, and a mass $M_{500}$=(2.3$\pm$0.5)$\times$10$^{14}$ M$_{\sun}$, corresponding 
 to a total mass $M_{200}$ $\sim 3.5 \times$10$^{14}$ M$_{\sun}$. 

\subsection{Optical/NIR photometry}

The initial follow-up procedure of the XDCP candidates is to perform 2-band imaging in 
the NIR and optical, using SOFI (Moorwood et al. \cite{moor}) and EMMI (Dekker et al. \cite{dekker}) 
mounted on the 3.5m ESO/NTT telescope.
In 2007 (Prog ID 079.A-0634(B), PI H. Quintana) we acquired 1h of H-band imaging with SOFI
using the \textit{Large field} mode, corresponding to a 5$\arcmin \times$5 $\arcmin$ FoV, and
 a pixel scale of 0.288$\arcsec$/pixel. 
I-band imaging (30 min) was taken with EMMI in the RILD (Red Imaging and Low Dispersion 
Spectroscopy) mode, which is suitable for observations redder than 400 nm, 
using the I\#610 filter. In this mode the FoV is 9.1$\arcmin \times$ 9.9$\arcmin$, the pixel scale is 0.335 
$\arcsec$/pix, and we used a binning of 2$\times$2. 
The H-band data was reduced with the package ESO/MVM (Vandame \cite{vandame}) and 
the I-band imaging was processed with IRAF routines. 
The seeing of the images is on average 0.8$\arcsec$ in the H-band image, and 1.0$\arcsec$ in the I-band.
The photometric zero points are 22.52$\pm$0.02 mag and 25.61$\pm$0.01 mag in the H- and I-band, 
respectively, based on 5 standard stars taken in the nights of the science observations.

Pre-imaging in the z-band (10 min) was obtained in 2008 for the upcoming spectroscopic 
observations, using VLT/FORS2 (Prog ID 081.A-0312(A)), PI H. Quintana). 
In Fig. 1 we show the  colour composite (IzH), where we clearly identify an overdensity of red 
galaxies coincident with the X-ray emission. 

\subsection{Spectroscopic data}

In 2009 we acquired medium-deep observations (2.2 hours) with VLT/FORS2 (Prog ID 084.A-0844(B), PI H. Quintana),
to obtain spectroscopic redshifts for the galaxies associated with the cluster
We observed one MXU-mode slit-mask using the 300I grism that covers the wavelength range 
5800-10500 \AA, with a resolution of R=660. 
The data reduction was performed with an adapted version of the VIMOS 
Interactive Pipeline and Graphical Interface (VIPGI, Scodeggio et al. \cite{scodeggio}).

The target selection for the spectroscopic mask was based primarily on the initial I-H colour-magnitude diagram (CMD). 
Technical constraints concerning the positioning of the slits in the candidate cluster center limited the number of targets 
to five slits in the region encompassing the X-ray emission, and a total of seven slits within a cluster-centric distance of 
1$\arcmin$.
Three out of seven targeted galaxies within 46$\arcsec$ from the X-ray centroid have a redshift of $z\sim$1.58.  
The cluster redshift is thus $z$ = 1.579 $\pm$ 0.003. 
The redshift measurements were based on the detection of the [OII] ($\lambda$ 3727) emission line and the FeII 
absorption series (see Table 1 and Fig. 2), and by cross-correlating the spectra with a set of spectral templates.
 Two of the three confirmed members have [OII] emission, indicating ongoing star formation, 
with equivalent widths of (-32$\pm$6) \AA~and (-88$\pm$8) \AA~for galaxies ID 2 and ID 3 
respectively. 

\begin{SCtable*}
\caption{Spectroscopic redshifts of 7 galaxies located within 1$\arcmin$ of the X-ray centroid, including the 
three cluster members of XMMU\,J0044 (ID 1, 2, 3). We list total Vega H-band magnitudes, I$-$H colours, 
projected distances in arcseconds relative to the 
X-ray center, spectroscopic redshift and spectral features. }
{\small
\centering
\begin{tabular}{ |c |c |c |c |c |c |c |c|}
\hline \hline

ID & RA      & DEC      & H        & I$-$H    & d$_{\mathrm{c}}$   & $z_{\mathrm{spec}}$   &  Features \\

     &  J2000 &  J2000  &  mag  & mag       &  \arcsec                            &                           &    \\

\hline
1  & 00:44:05.63 &  -20:33:53.99 &  18.6  &   3.07   &  8     & 1.5790$\pm$0.0003  &  FeII, MgII  \\    
2  & 00:44:05.33 & -20:33:13.72  &  19.8  &   3.23  &  46    & 1.5788$\pm$0.0003  &  [OII], FeII   \\    
3  & 00:44:05.57 & -20:34:16.10  &  19.3  &   2.27  &  17   & 1.5721$\pm$0.0008  &   [OII]           \\       
\hline
4  & 00:44:04.57 & -20:34:10.76  &  19.7  & 3.17      &  14    & 0.5924$\pm$0.0002  &    [OII], H$_{\beta}$, [OIII]    \\  
5  & 00:44:04.40 & -20:33:42.05  &  18.9  & 4.29      &   21   & 1.0709$\pm$0.0009  &    CaH/K                \\  
6  & 00:44:02.73 & -20:34:02.19  &  20.1  & 3.79      &  35   & 1.3271$\pm$0.0008  &     CaH/K                \\  
7  & 00:44:02.41 & -20:33:46.67  &  22.1  & 2.02     &  41   & 1.0178$\pm$0.0007  &                   [OII]            \\  
\hline
\end{tabular}
}
\end{SCtable*}

\subsection{Colour-magnitude diagram}

In order to identify a sequence of red elliptical galaxies that typically characterize relaxed galaxy 
clusters (Gladders \& Yee \cite{gladders}), we investigated the I-H colour-magnitude diagram 
using the cluster redshift obtained in the previous section. 

We applied the IRAF task GAUSS to perform the PSF matching of the two bands and
used SExtractor (Bertin \& Arnouts \cite{bertin}) in \textit{dual image mode} to produce the photometric catalogs.
The CMD is presented in Fig. 2-left panel, showing only objects within a cluster-centric distance of 1$\arcmin$.
 To limit intrinsic galaxy colour gradients, the I-H colour was measured 
in small apertures of 1.3$\arcsec$ diameter, which is slightly larger than the PSF.  
Errors on the colours are derived from the SExtractor I- and H-band uncertainties, ranging from 0.06 to 0.49 mag (for the faint, red galaxies).
Total H-band magnitudes were obtained with the SExtractor parameter {\tt MAGAUTO}.
The galactic extinction was accounted for using dust extinction maps
from Schlegel et al. (\cite{schlegel}) available at the Nasa Extragalactic Database. We
retrieved E(B-V) = 0.018 mag, corresponding to corrections in the I- and H-bands
of 0.034 mag and 0.010 mag, respectively.

We estimate $m$* to be at H=20.1 mag (vertical dotted line in Fig. 2) using the evolution of 
$m$*$\_$H predicted by Kodama \& Arimoto (\cite{kodama}) models with $z_{f}$=3. 
We note that these expectations are confirmed up to $z$=1.4, with the study of the 
H-band luminosity function of XMMUJ2235 (Strazzullo et al. \cite{veronica}).
We identify a sequence of 14 red galaxies located within a radius 
of 1$\arcmin$ from the X-ray center (Fig. 1, left-panel), by applying a reasonable colour-cut of 3.7 $<$ I-H $<$ 4.6, 
and considering only galaxies up to the 50\% completeness limit in the H-band.

In addition to the three cluster members and the four interlopers, 
we also indicate the colour of the large, central galaxy.
The three cluster members have I-H colours that are bluer than what one would naively expect 
using Simple Stellar Population (SSP) models considering a Salpeter IMF (Salpeter \cite{salpeter}) with solar metallicity
of passively evolving ellipticals with formation redshift $z_{f}$=3 and 5, 
that predict I-H equal to 3.75 and 4.0, respectively, at the cluster redshift.
In particular, galaxy ID 3 has a very prominent [OII] line, which is reflected in its I-H colour, 1 magnitude bluer 
than the other two spectroscopic members. This rather blue galaxy is located in the core, which is unusual 
in the X-ray clusters studied so far out to $z\sim$1.4, since star forming galaxies in clusters are typically located 
in the outskirts of the cluster. The confirmed members and the large central galaxy show signs of distorted 
morphology and resemble late-type galaxies. Sill, only with the \textit{Hubble Space Telescope} it will be 
possible to accurately assess the structure of these galaxies.

\begin{figure*}
\begin{center}
\includegraphics[width=6.8cm,angle=0]{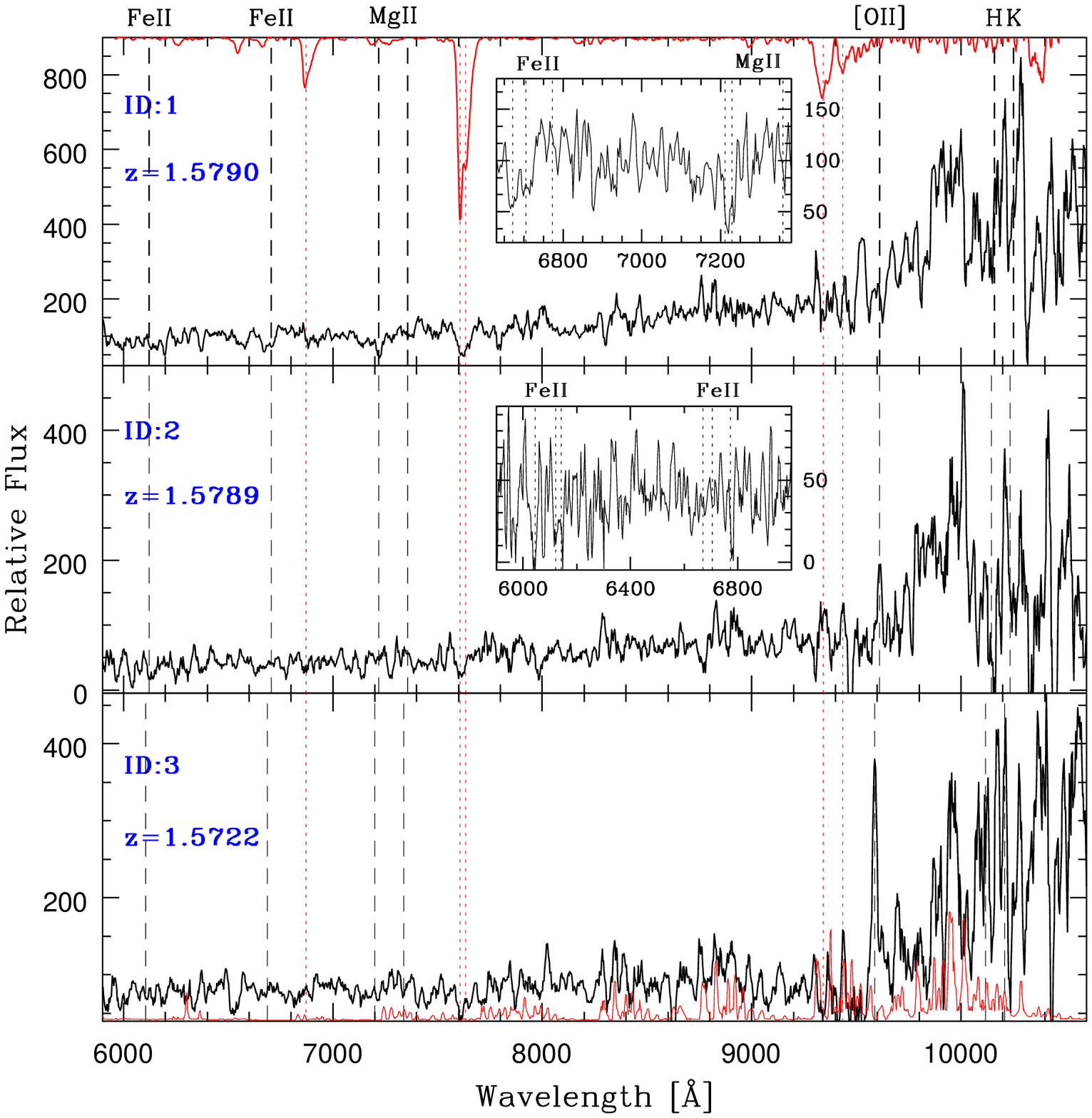}
\includegraphics[width=9.5cm,angle=0]{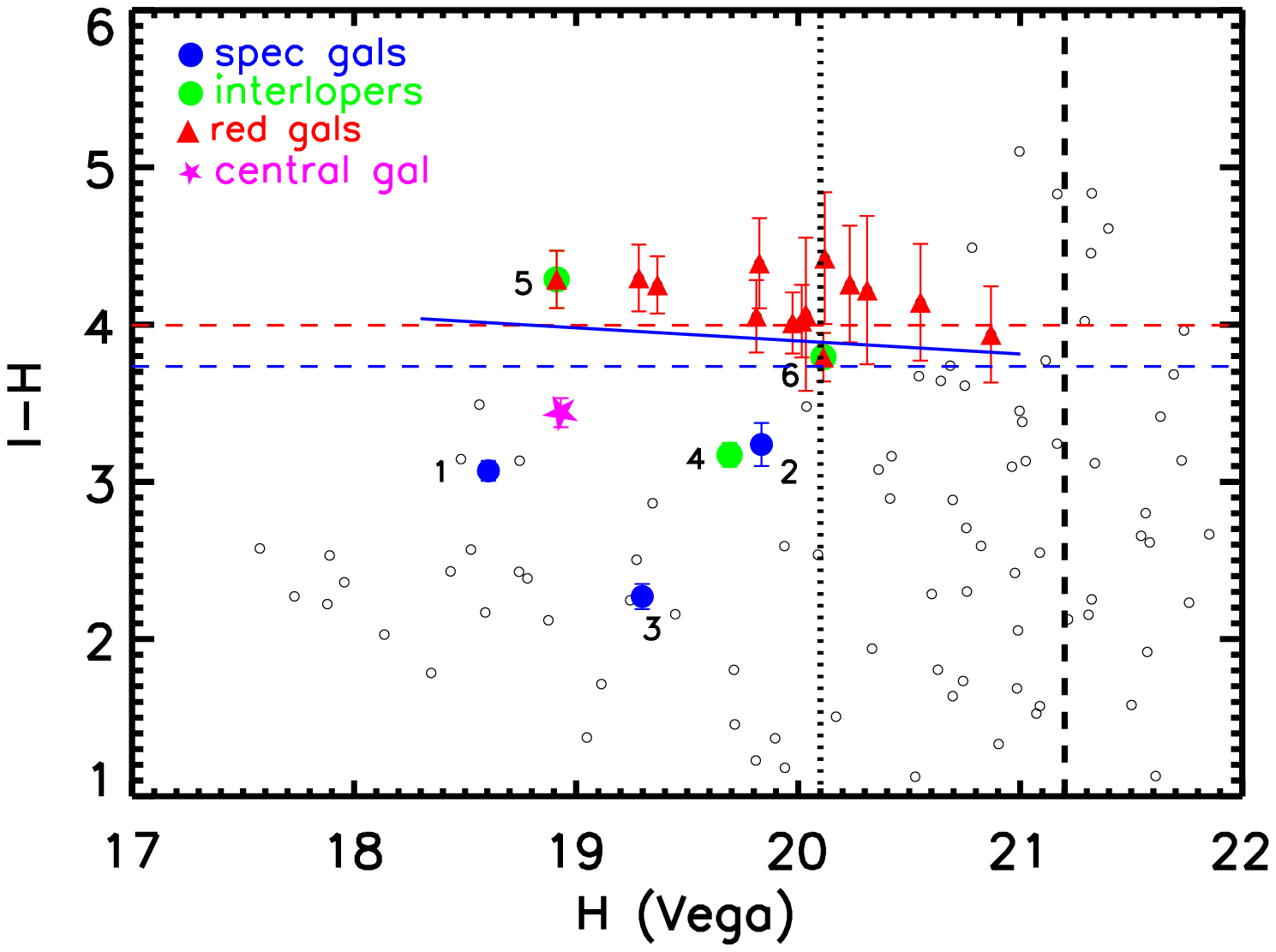}
\end{center}
\vspace*{-0.5cm}
\caption{\textit{Left:} FORS2 spectra of the three confirmed cluster members, smoothed with a 7 pixel boxcar filter. 
In the first panel we overlay the telluric absorption spectrum in red. Sky emission lines are shown in red 
in the bottom panel.  The inset shows the FeII and MgII lines.
\textit{Left:} I-H colour-magnitude diagram of XMMUJ0044. Only objects within 1$\arcmin$ radial distance from the cluster 
X -ray center are shown. The three confirmed members are shown in blue circles and the interlopers are marked in green. 
The ID numbers refer to Table 1. The large, central galaxy is identified by a magenta star. 
We also flag the red (3.7 $<$ I-H $<$ 4.6) galaxies that are likely associated with the 
cluster (red triangles). The horizontal dashed lines refer to SSP model galaxies with different stellar formation epochs 
at the cluster redshift (red line: $z_{f}$=5, blue line: $z_{f}$=3). The solid blue line refers to the de-evolved CMD of 
XMMUJ2235 at $z$=1.58. The 50\% completeness limit in the H-band is shown by a vertical dashed line.
}
 \label{spec}
\end{figure*}

\section{Discussion}

The X-ray analysis, based on 110 cluster counts in the [0.35-2.4] keV band, allowed us to 
estimate a bolometric luminosity of $\sim$6$\times 10^{44}$ erg/s.
This value is typical of low-$z$ massive galaxy clusters and corresponds to about three times the luminosity of 
another cluster at a very similar redshift, XMMUJ1007.4+1237 at $z$=1.56, which has
 $L_{bol} (r<R_{500}) \sim 2.1\times10^{44}$ erg/s   (Fassbender et al. \cite{fassben11}). 
Furthermore, XMMUJ0044 is also significantly more luminous than the more distant poor cluster/groups discovered 
at $z$=1.62, CIG2018.3-0510 with $L_{[0.1-0.24]} \sim 0.3\times10^{44}$ erg/s  ($T_{est}$=1.7 keV, Tanaka et al. \cite{tanaka})  
and CLJ1449+0856 at $z$=2.07 with $L_{[0.1-0.24]} \sim 0.7\times10^{44}$ erg/s ($T_{est}$=2.0 keV, Gobat et al. \cite{gobat}).
We note that recent work using shallow \textit{Chandra} data of \textit{Spitzer}/IRAC cluster ISCS J1438.1+3414 
at $z$=1.49, shows this to be a massive cluster, with $L_{[0.5-2.0]}$=1.0$\times10^{44}$ erg/s and luminosity derived 
$M_{200}\sim2.2\times10^{14} M_{\odot}$ (Brodwin et al. \cite{brodwin}).

The optical/NIR data depicts a population of red galaxies centered on the X-ray contours. 
In addition to the SSP models, we also compared the sequence 
of 14 red galaxies with what we would expect from a de-evolved CMD of XMMUJ2235 at $z$=1.58 (solid blue line in 
Fig. 2). This empirical expectation is in excellent agreement with predictions from Kodama \& Arimoto (\cite{kodama}) 
models for $z_{f}$=5, therefore the observed red-sequence (RS) of XMMUJ0044 appears fainter and redder than expected. 
A possible explanation is that this sequence is not entirely formed by normal passive galaxies, but instead it is 
populated by dusty star-forming galaxies (Pierini et al. 2005). On the other hand, the large uncertainty in the colours 
of the faint, red galaxies prevents a more accurate assessment of the locus of the red-sequence.
The bright population in the cluster core (including ID 1 and ID 3) is not dominated by passive galaxies as 
observed at lower redshifts, instead we observe bright blue star forming galaxies in the center of the cluster 
that might dominate the bright end of the colour-magnitude diagram. 
Similar findings have been reported at $z$=1.45 (Hilton et al. \cite{hilton}) and $z$=1.6 (Tran et al. \cite{tran}).

The central group of three red galaxies for which we have no spectroscopic information is of particular interest.
The I-H colour of the largest and brightest galaxy (magenta star in Fig. 2) is redder (by 0.26 mag) than the brightest 
spectroscopic member (ID 1, only 2$\arcsec$ away). The other two smaller central galaxies are about 1 
magnitude fainter relative to the former and are significantly redder (I-H=4.13 and 4.17 mag). 
We may argue that this group is in the process of merging that will eventually lead to the formation 
of the brightest cluster galaxy.

\section{Conclusions}

In this letter we present a multi-wavelength analysis of an X-ray luminous galaxy cluster at $z$=1.579, 
XMMUJ0044.0-2033, that was serendipitously discovered in the archive of the XMM-\textit{Newton} observatory. 
Here we summarize our main results:

\begin{itemize}

\item XMMUJ0044 was detected as a bright, compact but significantly extended source in the XMM-\textit{Newton} data;

\item in dedicated observations in the H- and I-bands we found an overdensity of red galaxies 
coincident with the X-ray peak; 

\item using FORS2 spectroscopy we secured three cluster members 
within $r$=46$\arcsec$ from the cluster X-ray peak, with $z\sim$1.58; 

\item the I-H colour-magnitude diagram shows that the three confirmed members have 
 bluer colours with respect to the sequence of 14 galaxies with [3.7 $<$ I-H $<$ 4.6].
The cluster members show a distorted morphology and two of them exhibit [OII] line emission; 

\item we identified an interesting central group of 3 galaxies located at $\sim 2 \arcsec$ from the 
cluster X-ray centroid. The brightest galaxy is redder than the confirmed members, but bluer
than the two smaller companions, which sit on the observed RS;

\item knowing the cluster redshift, we measured a luminosity, $L_{bol, 40\arcsec}$=5.8$\times$10$^{44}$ erg/s 
or $L_{bol, R_{500}}$=6.1$\times$10$^{44}$ erg/s, and estimated a range of the cluster mass, 
$M_{tot}\sim$ 3.5-5.0$\times$10$^{14} M_{\sun}$, depending on the methods used to scale the luminosity. 

\end{itemize}

The analysis presented here confirms XMMUJ0044 as one of the most 
massive, distant clusters known to date, characterized by a high X-ray luminosity.
Deeper, high-resolution X-ray data will allow us to measure the ICM temperature, which is crucial to better assess the 
dynamical state of this cluster, as well as the gas metal content. 
Additional spectroscopy and already awarded HAWK-I J/Ks imaging data will enable a more complete 
characterization of the galaxy cluster population.

\acknowledgements

JSS thanks Simona Mei for fruitful discussions and advice on the CMD and Gabriel Pratt for comments on the X-ray analysis.
This work was supported by the DFG under grants Schw 536/24-2, BO
702/16-3, and the DLR under grant 50 QR 0802. 
This research has made use of the NASA/IPAC Extragalactic Data base
(NED) which is operated by the Jet Propulsion Laboratory, California Institute
of Technology.

\bibliographystyle{aa}

\end{document}